\documentstyle[aps,epsf]{revtex}  

\begin{document}        
\draft
\newcommand{\beq}{\begin{equation}}
\newcommand{\dd}{\partial}
\newcommand{\eeq}{\end{equation}}
\newcommand{\bea}{\begin{eqnarray}}
\newcommand{\eea}{\end{eqnarray}}
\newcommand{\nue}{\nu_e}
\newcommand{\nueb}{\bar{\nu}_e}
\newcommand{\numu}{\nu_\mu}
\newcommand{\numub}{\bar{\nu}_\mu}
\newcommand{\nutau}{\nu_\tau}
\newcommand{\nutaub}{\bar{\nu}_\tau}
\newcommand{\nus}{\nu_s}
\newcommand{\nusb}{\bar{\nu}_s}
\newcommand{\numt}{\nu_{\mu,\tau}}
\newcommand{\numtb}{\bar{\nu}_{\mu,\tau}}
\baselineskip 14pt
\title{Explanations of pulsar velocities} 
\author{Alexander Kusenko}
\address{Department of Physics and Astronomy, UCLA, Los Angeles, CA
90095-1547 }
%
\maketitle              

\begin{abstract}        

     Several mechanisms based on neutrino oscillations can explain the
observed motions of pulsars if the magnetic field in their interiors 
is of order $10^{14}-10^{15}$~G.

\
\end{abstract}   	

\pacs{UCLA/99/TEP/9}

\section{Introduction}               

The proper motions of pulsars~\cite{astro} present an intriguing
astrophysical puzzle.  The measured velocities of pulsars exceed those of
the ordinary stars in the galaxy by at least an order of magnitude.  The
data suggest that neutron stars receive a powerful ``kick'' at birth.
Whatever the cause of the kick, the same mechanism may also explain the
rotations of pulsars under some conditions~\cite{sp}.

The origin of the birth velocities is unclear.  Born in a supernova
explosion, a pulsar may receive a substantial kick due to the
asymmetries in the collapse, explosion, and the neutrino emission affected
by convection~\cite{explosion}.  Evolution of close binary systems may also
produce rapidly moving pulsars~\cite{b}.  It was also suggested~\cite{hd}
that the pulsar may be accelerated during the first few months after the
supernova explosion by its electromagnetic radiation, the asymmetry
resulting from the magnetic dipole moment being inclined to the rotation
axis and offset from the center of the star.  Most of these mechanisms,
however, have difficulties explaining the magnitudes of pulsar spatial
velocities in excess of 100~km/s.  Although the average pulsar velocity is
only a factor of a few higher, there is a substantial population of pulsars
which move faster than 700~km/s, some as fast as 1000~km/s~\cite{astro}. 

Neutrinos carry away most of the energy, $\sim 10^{53}$~erg, of the
supernova explosion.  A 1\% asymmetry in the distribution of the neutrino
momenta is sufficient to explain the pulsar ``kicks''.  A strong magnetic
field inside the neutron star could set the preferred direction.  However,
the neutrino interactions with the magnetic field are hopelessly weak.

Ordinary electroweak processes~\cite{chugai} cannot account for the
necessary anisotropy of the neutrino emission~\cite{vil}.  The possibility
of a cumulative build-up of the asymmetry due to some parity-violating
scattering effects has also been considered~\cite{cumulative}.  However, in
statistical equilibrium, the asymmetry does not build up even if the
scattering amplitudes are asymmetric~\cite{vil,eq}.  Although some net
asymmetry develops because of the departure from equilibrium, it is too
small to explain the pulsar velocities for realistic values of the magnetic
field inside the neutron star~\cite{vil,al}.

There is a class of mechanisms, however, that can explain the birth
velocities of pulsars as long as the magnetic field inside a neutron star
is $10^{14} - 10^{15}$~G.  These mechanisms~\cite{ks96,ks97,ks98,others}
have some common features.  First, the conversions of some neutrino $\nu$
into a different type of neutrino, $\nu'$, occurs when one of these
neutrinos is free-streaming while the other one is not.  The free-streaming
component is out of equilibrium with the rest of the star, which prevents
the wash-out of the asymmetry.  Second, the position of the transition
point it affected by the magnetic field.  I will review two possible
explanations, which do not require any exotic neutrino interactions and
rely only on the established neutrino properties, namely matter-enhanced
neutrino oscillations.  The additional assumptions about the existence of
sterile neutrinos~\cite{rm} and the neutrino masses appear plausible from
the point of view of particle physics.

\section{Pulsar kicks from neutrino oscillations }

As neutrinos pass through matter, they experience an effective potential

\bea
V(\nus) & = & 0  \label{Vnus} \\
V(\nue)& = & -V(\nueb) =  V_0 \: (3 \, Y_e-1+4 \, Y_{\nue}) \label{Vnue} \\
V(\nu_{\mu,\tau}) & = & -V(\bar{\nu}_{\mu,\tau}) = V_0 \: ( Y_e-1+2 \, 
Y_{\nue}) \ 
+\frac{e G_{_F}}{\sqrt{2}} \left ( \frac{3 N_e}{\pi^4} 
\right )^{1/3}
\frac{\vec{k} \cdot \vec{B}}{|\vec{k}|} \label{Vnumu}
\eea
where $Y_e$ ($Y_{\nue}$) is the ratio of the number density of electrons
(neutrinos) to that of neutrons, $\vec{B}$ is the magnetic field, 
$\vec{k}$ is the neutrino momentum, $V_0=10 \: \rm{eV} \: (\rho/10^{14} g
\, cm^{-3} )$.  The magnetic field dependent term in equation (\ref{Vnumu})
arises from a one-loop finite-density contribution \cite{magn,smirnov} 
to the self-energy of a neutrino propagating in a magnetized 
medium.  An excellent review of the neutrino ``refraction'' in
magnetized medium is found in Ref. \cite{smirnov}.

The condition for resonant~\cite{msw} oscillation $\nu_i \leftrightarrow
\nu_j$ is

\beq
\frac{m_i^2}{2 k} \: cos \, 2\theta_{ij} + V(\nu_i) = 
\frac{m_j^2}{2 k} \: cos \, 2\theta_{ij} + V(\nu_j)  
\label{res}
\eeq
where $\nu_{i,j}$ can be either a neutrino or an anti-neutrino. 

The neutron star can receive a kick if the following two conditions
\cite{ks96,ks97,ks98} are satisfied: (1)~the
adiabatic\footnote{Non-adiabatic oscillations are discussed in
Ref.~\cite{kim}} oscillation $\nu_i \leftrightarrow \nu_j$ occurs at a
point inside the $i$-neutrinosphere but outside the $j$-neutrinosphere; and
(2)~the difference $[V(\nu_i)-V(\nu_j)]$ contains a piece that depends on
the relative orientation of the magnetic field $\vec{B}$ and the momentum
of the outgoing neutrinos, $\vec{k}$.  If the first condition is satisfied,
the effective neutrinosphere of $\nu_j$ coincides with the surface formed
by the points of resonance.  The second condition ensures that this surface
(a ``resonance-sphere'') is deformed by the magnetic field in such a way
that it will be further from the center of the star when $(\vec{k} \cdot
\vec{B}) > 0$, and nearer when $(\vec{k} \cdot \vec{B})<0$.  The average
momentum carried away by the neutrinos depends on the temperature of the
region from which they exit.  The deeper inside the star, the higher is the
temperature during the neutrino cooling phase.  Therefore, neutrinos coming
out in different directions carry momenta which depend on the relative
orientation of $\vec{k}$ and $\vec{B}$.  This causes the asymmetry in the
momentum distribution.  An $1\%$ asymmetry is sufficient to generate birth
velocities of pulsars consistent with observation.

Let us use two different models for the neutrino emission to calculate the
kick from the active-sterile and the active neutrinos, respectively.  As
shown in Ref.~\cite{ks98}, these two models are in good agreement. 

\section{Oscillations into sterile neutrinos
}

Since the sterile neutrinos have a zero-radius neutrinosphere, $\nu_s
\leftrightarrow \numtb$ oscillations can be the cause of the pulsar motions
if $m(\nus) > m(\numt)$.  If, on the other hand, $m(\nus) < m(\numt)$,
$\nu_s \leftrightarrow \numt$ oscillations can play the same role.  

\begin{figure}	
\centerline{\epsfxsize 2.5 truein \epsfbox{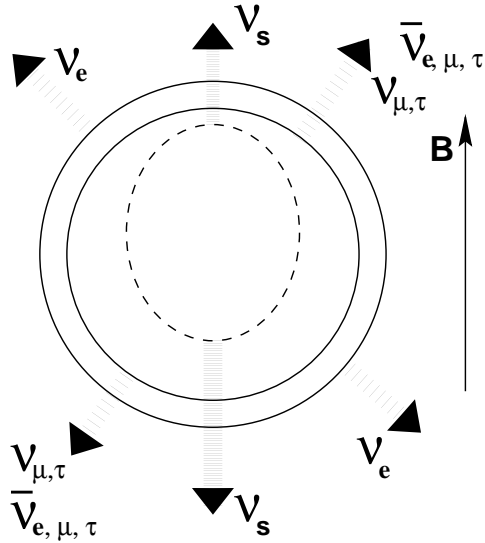}}   
\vskip .2 cm
\caption[]{
\label{sterile}
\small Sterile neutrinos produced by oscillations are emitted from regions
of different temperatures inside a neutron star.}
\end{figure}

In the presence of the magnetic field, the condition (\ref{res}) is
satisfied at different distances $r$ from the center (Fig.~\ref{sterile}),
depending on the value of the $(\vec{k} \cdot \vec{B})$ term in
(\ref{res}). The surface of the resonance is, therefore,

\begin{equation}
r(\phi) = r_0 + \delta \: cos \, \phi, 
\end{equation}
where $cos \, \phi= (\vec{k} \cdot \vec{B})/k$ and $\delta$ is determined
by the equation 
$(d N_n(r)/dr) \delta \approx 
e \left ( 3 N_e/\pi^4 \right )^{1/3} B$.
This yields~\cite{ks97} 

\begin{equation}
\delta = 
\frac{e \mu_e}{ \pi^2} \: B \left / \frac{dN_n(r)}{dr} \right. ,
\label{delta}
\end{equation}
where $\mu_e \approx (3 \pi^2 N_e)^{1/3} $ is the chemical potential of the
degenerate (relativistic)  electron gas.

Assuming a black-body radiation luminosity $\propto T^4$ for the effective
neutrinosphere, the asymmetry in momentum distribution~\cite{ks97} is  
\begin{equation}
\frac{\Delta k}{k} = \frac{4 e}{3 \pi^2} \: \left ( \frac{\mu_e}{T}
\frac{dT}{dN_n} \right) B, 
\label{dk1}
\end{equation}

To calculate the derivative in (\ref{dk1}), we use the relation between the
density and the temperature of a non-relativistic Fermi gas. Finally, 
\begin{equation}
\frac{\Delta k}{k} = \frac{4 e\sqrt{2}}{\pi^2} \: 
\frac{\mu_e \mu_n^{1/2}}{m_n^{3/2}T^2} \ B =
0.01 \frac{B}{3\times 10^{15} {\rm G}} 
\label{dk2}
\end{equation}
if the neutrino oscillations take place in the core of the neutron star, at
density of order $10^{14} \, {\rm g\,cm^{-3}}$.  The neutrino oscillations
take place at such a high density if one of the neutrinos has mass in the
keV range, while the other one is much lighter.

The mixing angle can be very small, because the adiabaticity
condition is satisfied if 

\begin{equation}
l_{osc} \approx \left (\frac{1}{2\pi} \ 
\frac{\Delta m^2}{2 k} \ sin \, 2 \theta
\right )^{-1} \approx \frac{10^{-2} \: {\rm cm}}{sin \, 2 \theta }
\end{equation}
is smaller than the typical scale of the density variations.  
Thus the oscillations remain adiabatic as long as $sin^2 \, 2 \theta > 
10^{-8}$.

\section{Oscillations of active neutrinos}

The active neutrino oscillations can also explain the pulsar
kick~\cite{ks96}.  The magnitude of the kick can be calculated using a
model for neutrino transfer used in the previous section~\cite{ks96}.  That
is, one can assume that the neutrinos are emitted from a ``hard''
neutrinosphere with temperature $T(r)$ and that their energies are
described by the Stefan-Boltzmann law.  Alternatively, we can use the
Eddington model for the atmosphere which was used by Schinder and
Shapiro~\cite{ss} to describe the emission of a single neutrino species.
One can generalize it to include several types of neutrinos.\footnote{A
recent attempt~\cite{jr} to use the Eddington model for the neutrino
transfer failed to produce a correct result because the neutrino absorption
$\nu_e n \rightarrow e^-p^+$ was neglected, and also because the different
neutrino opacities were assumed to be equal to each other.  The
assumption~\cite{jr} that the effect of neutrino oscillations can be
accounted for in a simplistic model with one neutrino species and a
deformed core-atmosphere boundary is also incorrect because the temperature
profile is determined by the emission of six neutrino types, five of which
are emitted isotropically.  The neutrinos of the sixth flavor, which have
an anisotropic momentum distribution, cause negligible (down by at least a
factor of 6) asymmetry in the temperature profile.  When the neutrino
absorption is included, the Eddington model gives the same result for the
kick~\cite{ks98} as the model with ``hard
neutrinospheres''~\cite{ks96,ks97}.}

In the diffusion approximation, the distribution functions $f$ are taken in
the form~\cite{ss}:   
\begin{equation}
f_{\nu_i} \approx f_{\bar{\nu}_i}
\approx f^{eq} + \frac{\xi}{\Lambda_i} \frac{\partial
f^{eq}}{\partial m}, 
\label{f_diff}
\end{equation}
where $f^{eq}$ is the distribution function in equilibrium, $\Lambda_i$
denote the respective opacities, $m$ is the column mass density, $m=\int
\rho \: dx$, $\xi=cos \alpha$, and $\alpha $ is the normal angle of 
the neutrino velocity to the surface. 
At the surface, one imposes the same boundary condition for all the
distribution functions, namely 
\begin{equation}
f_{\nu_i}(m,\xi)= \left \{ 
\begin{array}{ll}
0, &  {\rm for} \ \xi < 0, \\
2 f^{eq}, &  {\rm for} \ \xi > 0.
\end{array} \right.
\label{bc}
\end{equation}
However, the differences in $\Lambda_i$ produce the unequal distributions
for different neutrino types.

Generalizing the discussion of Refs.~\cite{ss} to include six flavors,
three neutrinos and three antineutrinos, one can write the energy flux as 

\begin{equation}
F=2\pi \int_0^\infty E^3 dE \int_{-1}^1 \xi d \xi \ 
\sum_{i=1}^{3} (f_{\nu_i}+f_{\bar{\nu}_i}),
\end{equation}
We will assume that $\Lambda_i=\Lambda_i^{(0)} (E^2/E^2_0)$.  

\begin{figure}	
\centerline{\epsfxsize 2.5 truein \epsfbox{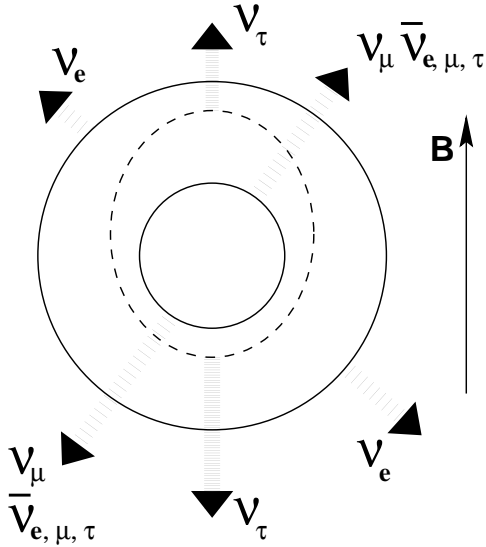}}
\vskip .2 cm
\caption[]{
\label{active}
\small  Oscillations {$ \nu_\tau \leftrightarrow \nu_e $} result in
anisotropic emission of {$\tau $}-neutrinos. The upward-going
{$\nu_\tau$}'s escape from the region with a lower temperature than that of
the downward-going {$\tau $}-neutrinos.  
}
\end{figure}

We use the expressions for $f_{\nu_i}$ from equation (\ref{f_diff}). 
Changing the order of differentiation with
respect to $m$ and integration over $E$ and $\xi$, and 
using the fact that $f^{eq}$ is isotropic, we arrive at
the result similar to that of Ref.~\cite{ss}: 

\begin{equation}
F=\frac{2\pi^3}{9} E_0^2  \left [ \sum_{i=1}^{3} \frac{2}{\Lambda^{(0)}_i} 
\right ] 
\frac{\partial T^2}{\partial m}.
\end{equation}

The basic assumption of the model is that flux $F$ is conserved.  In other
words, the neutrino absorptions   $\nu_e n \rightarrow e^- p^+$ are
neglected.  Since the sum in brackets, as well as the flux $F$ are
treated~\cite{ss} as constants with respect to $m$, one can solve
for $T^2$:  

\begin{equation}
T^2(m) = \frac{9}{2\pi^3} E_0^{-2} \left [ \sum_{i=1}^{3}
\frac{2}{\Lambda^{(0)}_i}  \right ]^{-1} F \, m+ \left(\frac{30}{7\pi^5} F
\right )^{1/2} 
\label{t2}
\end{equation}

Swapping the two flavors in equation (\ref{t2}) leaves the
temperature unchanged in the Eddington approximation.  Hence, neutrino
oscillations do not alter the temperature profile in this
approximation.

We will now include the absorptions of neutrinos. 

Some of the electron neutrinos are absorbed on their passage through the
atmosphere thanks to the charged-current process

\begin{equation}
\nu_e n \rightarrow e^- p^+.
\end{equation}
The cross section for this reaction is $\sigma = 1.8 \: G_{_F}^2
E_\nu^2 $, where $E_\nu$ is the neutrino energy.  The total momentum
transfered to the neutron star by the passing neutrinos depends on the
energy. 

Both numerical and analytical calculations show that the muon and tau
neutrinos leaving the core have much higher mean energies than the electron
neutrinos~\cite{suzuki,energies}.  Below the point of MSW~\cite{msw}
resonance the electron neutrinos have the mean energies $\approx 10$~MeV,
while the muon and tau neutrinos have energies $\approx 25$~MeV.

The origin of the kick in this description is that the neutrinos spend more
time as energetic electron neutrinos on one side of the star than on the other
side, hence creating the asymmetry.  Although the temperature profile
remains unchanged in Eddington approximation, the unequal numbers of
neutrino absorptions push the star, so that the total momentum is
conserved. 

Below the resonance
$E_{\nu_e}<E_{\nu_{\tau,\mu}}$. Above the resonance, this relation is
inverted.  The energy deposition into the nuclear matter depends on the
distance the electron neutrino has traveled with a higher energy.  This
distance is affected by the direction of the magnetic field relative to the
neutrino momentum.

We assume that the resonant conversion $\nu_e \leftrightarrow \nu_\tau$
takes place at the point $r=r_0+\delta(\phi); \delta (\phi)=\delta_0 \cos
\phi$.  The position of the resonance depends on the
magnetic field $B$ inside the star~\cite{ks96}: 

\begin{equation}
\delta_0=\frac{e \mu_e B}{2 \pi^2} \left / \frac{dN_e}{dr} \right., 
\end{equation}
where $N_e \approx Y_e N_n$ is the electron density and $\mu_e$ is the
electron chemical potential.

Below the resonance the  
$\tau$ neutrinos are more energetic than the electron neutrinos.  The
oscillations exchange the neutrino flavors, so that above  the resonance
the electron neutrinos are more energetic than the $\tau$ neutrinos.  The
number of neutrino absorptions in the layer of thickness $2 \delta (\phi)$
around $r_0$ depends on the angle $\phi$ between the neutrino momentum and
the direction of the magnetic field.  Each occurrence of the neutrino
absorption transfers the momentum $E_{\nu_e}$ to the star.  The difference
in the numbers of collisions per electron neutrino between the directions
$\phi$ and $\pi+\phi$ is  

\begin{eqnarray}
\Delta k_e/E_{\nu_e} & = & 2 \: \delta(\phi) \: 
N_n \: [\sigma(E_1)-\sigma(E_2)] \\
& = & 1.8 \: G_{_F}^2 [E_1^2-E_2^2] \: \frac{\mu_e}{Y_e} \: \frac{eB}{\pi^2} 
\: h_{N_e} \:  \cos \phi,
\end{eqnarray}
where $h_{N_e}=[d(\ln N_e)/dr]^{-1}$. 

We use $Y_e\approx 0.1$, $E_1\approx 25$~MeV, $E_2\approx10$~MeV,
$\mu_e\approx 50$~MeV, and $h_{N_e}\approx 6$~km. 
After integrating over angles and taking into account that only one
neutrino species undergoes the conversion, we obtain the final result for the
asymmetry in the momentum deposited by the neutrinos: 

\begin{equation}
\frac{\Delta k}{k} = 0.01 \frac{B}{2\times 10^{14} {\rm G}},
\label{final} 
\end{equation}
which agrees with the estimates\footnote{We note in passing that we
estimated the kick in Refs.~\cite{ks96,ks97} assuming $\mu_e \approx {\rm
const}$. A different approximation, $Y_e \approx {\rm const}$, gives a
somewhat higher prediction for the magnitude of the magnetic
field~\cite{comment}.}~\cite{ks96,comment} that use a different model for
the neutrino emission.

Neutrinos also lose energy by scattering off the electrons.  Since the
electrons are degenerate, the final-state electron must have energy greater
than $\mu_e$.  Therefore, electron neutrinos lose from $0.2$ to $0.5$ of
their energy per collision in the neutrino-electron scattering.  However,
since $N_e \ll N_n $, this process can be neglected. 

One may worry whether the asymmetric absorption can produce some
back-reaction and change the temperature distribution inside the star 
altering our result (\ref{final}).  If such effect exists, it is 
beyond the scope of Eddington approximation, as is clear from equation
(\ref{t2}).  The only effect of the asymmetric absorption is to
make the star itself move, in accordance with the momentum conservation.
This is the origin of the kick (\ref{final}). 

Of course, in reality the back-reaction is not exactly zero.  The most
serious drawback of Eddington model, pointed out in Ref.~\cite{ss}, is that
diffusion approximation breaks down in the region of interest, where the
neutrinos are weakly interacting.  Another problem has to do with
inclusion of neutrino absorptions and neutrino oscillations~\cite{ss}. 
However, to the extent we believe this
approximation, the pulsar kick is given by equation~(\ref{final}). 

\section{Conclusion}

The neutrino oscillations can explain the motions of pulsars. 
Although many alternatives have been proposed, all of them fail
to explain the large magnitudes of the pulsar velocities.  

If the pulsar kick velocities are due to $\nu_e \leftrightarrow
\nu_{\mu,\tau} $ conversions, one of the neutrinos must have mass $\sim 
100$~eV (assuming small mixing) and must decay on the cosmological time
scales not to overclose the Universe~\cite{ks96}.  This has profound 
implications for particle physics hinting at 
the existence of Majorons~\cite{peccei} or other physics beyond the
Standard Model that can facilitate the neutrino decay. 

If the active-to-sterile neutrino oscillations~\cite{ks97} are responsible
for pulsar velocities, the prediction for the sterile neutrino to have a mass
of several keV is not in contradiction with any of the present bounds.  In
fact, the $\sim$keV mass sterile neutrino has been proposed as a
dark-matter candidate~\cite{fs}. 

Some other explanations~\cite{others} that utilize new hypothetical
neutrino properties, but use a similar mechanism for generating the
asymmetry, can also explain large pulsar velocities.

\section{Acknowledgements}

The author thanks E.~S.~Phinney and G.~Segr\`e for many interesting and
stimulating discussions.  This work was supported in part by the US
Department of Energy grant DE-FG03-91ER40662.


\end{document}